\documentclass[12pt, a4paper]{elsarticle}

\let\today\relax
\makeatletter
\def\ps@pprintTitle{%
	\let\@oddhead\@empty
	\let\@evenhead\@empty
	\def\@oddfoot{\footnotesize\itshape
		{} \hfill\today}%
	\let\@evenfoot\@oddfoot
}
\makeatother

\usepackage{amssymb}

\usepackage{lineno}

\usepackage[pdftex]{hyperref}
\usepackage{xcolor}

\usepackage{listing}
\usepackage{framed}
\usepackage{lipsum}
\usepackage{soul}

\usepackage{float}
\restylefloat{table}

\journal{SoftwareX}
\usepackage{graphicx}
\usepackage[caption=false]{subfig}

\begin{document}

\begin{frontmatter}



\title{\textit{coherent WaveBurst}, a pipeline for unmodeled gravitational-wave data analysis}



\author[1,2]{M.~Drago}
\author[3]{V.~Gayathri}
\author[3]{S.~Klimenko}
\author[4]{C.~Lazzaro}
\author[5]{E.~Milotti}
\author[3]{G.~Mitselmakher}
\author[3]{V.~Necula}
\author[3]{B.~O'Brian}
\author[6,7]{G.~A.~Prodi}
\author[8]{F.~Salemi\corref{cor1}}
\ead{francesco.salemi@unitn.it}
\author[3]{M.~Szczepanczyk}
\author[9]{S.~Tiwari}
\author[10]{V.~Tiwari}
\author[11]{G.~Vedovato}
\author[12]{I.~Yakushin}
\cortext[cor1]{Corresponding author}

\address[1]{Gran Sasso Science Institute, Via F. Crispi 7, I-67100, L'Aquila, Italy}
\address[2]{INFN, Laboratori Nazionali del Gran Sasso, I-67100 Assergi, Italy }%
\address[3]{University of Florida, Gainesville, FL 32611, USA }%
\address[4]{Universit\`a di Padova, Dipartimento di Fisica e Astronomia, I-35131 Padova, Italy }
\address[5]{Dipartimento di Fisica, Universit\`a di Trieste and INFN Sezione di Trieste, Via Valerio, 2, I-34127 Trieste, Italy }
\address[6]{Universit\`a di Trento, Dipartimento di Matematica, I-38123 Povo, Trento, Italy}%
\address[7]{INFN, TIFPA, I-38123 Povo, Trento, Italy}%
\address[8]{Universit\`a di Trento, Dipartimento di Fisica, I-38123 Povo, Trento, Italy}%
\address[9]{Physik-Institut, University of Zurich, Winterthurerstrasse 190, 8057 Zurich, Switzerland }
\address[10]{Cardiff University, Cardiff CF24 3AA, United Kingdom}
\address[11]{INFN, Sezione di Padova, I-35131 Padova, Italy }
\address[12]{LIGO Livingston Observatory, Livingston, Louisiana 70754, USA}

\begin{abstract}

\textit{coherent WaveBurst} (cWB) is a highly configurable pipeline designed to detect a broad range of gravitational-wave (GW) transients in the data of the worldwide network of GW detectors. 
The algorithmic core of cWB is a time-frequency analysis with the Wilson-Daubechies-Meyer wavelets aimed at the identification of GW events without prior knowledge of the signal waveform.  cWB has been in active development  since 2003 and it has been used to analyze all scientific data collected by the LIGO-Virgo detectors ever since.  On September 14, 2015, the cWB low-latency search detected the first gravitational-wave event, GW150914, a merger of two black holes.
In 2019, a public open-source version of cWB has been released with GPLv3 license.	


\end{abstract}

\begin{keyword}
gravitational waves \sep signal processing \sep wavelets



\end{keyword}

\end{frontmatter}

\section{Motivation and significance}
\label{sec:intro}

The first direct detection of gravitational waves was accomplished by the LIGO-Virgo Collaboration in 2015, almost one century after the initial prediction by Albert Einstein \cite{Einstein1916}. On September 14, 2015, at 09:50:45 UTC, the low-latency instance of \textit{coherent WaveBurst} (cWB) \cite{CWBwavelet:2004km,CWBlikenet:2005kmrm,Klimenko:2015ypf,Klimenko:2008fu,gwburst.gitlab.io,cwb-doc}, an unmodeled 
search pipeline for the prompt detection of generic gravitational-waves (GWs), identified and reconstructed a chirping signal (see also Fig. \ref{f:SIMexamples}) in the data from the first observing run of the advanced LIGO detectors located
in Hanford (WA) and Livingston (LA) \footnote{The advanced Virgo interferometer in Cascina, Italy joined the GW network later on, in
2017.} and reported it within three minutes of data acquisition \cite{Abbott:2016blz}. 
Follow-up analyses by the LIGO-Virgo Collaboration established that the signal (later named GW150914) was consistent with the merger of a binary black hole  (BBH). This historic detection was achieved thanks to the extreme sensitivity of the advanced GW detectors and the sophisticated signal-processing methods used to separate signals from noise and all the remaining systematics, such as those implemented within cWB. 
The cWB pipeline is designed to detect a wide class of gravitational-wave signals and reconstruct their waveforms with minimal assumptions on the source model. It also extracts additional properties such as bandwidth, duration, sky location, and polarization state.
cWB searches for GWs both in low-latency mode (with a latency of few minutes) and offline.
The low-latency reconstructed sky location \cite{Aasi:2013wya,PhysRevD.83.102001} can be promptly shared with the partner observatories \cite{emfollow} which search for coincident electromagnetic counterparts.  
As an example of an offline analysis, cWB has been used for targeted searches for core-collapse supernovae (CCSNe) \cite{Iess:2020yqj,PhysRevD.101.084002,Srivastava:2019fcb,Abbott:2018qee,Abbott:2016tdt} -- one of the next most anticipated GW sources, and one of the highest priority tasks for GW detectors. 
cWB is also well suited for the detection of compact binary coalescences (CBC): 
CBC sources formed from combinations of neutron stars (NS) and black holes (BH), are among the most efficient emitters of gravitational waves, and when observed in conjunction with electromagnetic or neutrino signals yield important astrophysical insights \cite{TheLIGOScientific:2017qsa,Abbott:2018wiz,Monitor:2017mdv,ANTARES:2017bia,LIGOScientific:2019eut}. 
The cWB algorithms are robust with respect to a variety of CBC features including  higher multipoles, high mass ratios, misaligned spins, eccentric orbits and possible deviations from  general relativity, that may create mismatches between signal waveforms and simulated CBC templates. Recently, cWB played a significant role both in the identification of higher multipoles for GW190814, an event associated with the coalescence of a binary system with the most unequal mass ratio yet measured with gravitational waves \cite{Abbott:2020khf} and in the detection of GW190521, the most massive and most distant black hole merger yet observed in gravitational waves, the first direct detection of an intermediate-mass black hole binary \cite{Abbott:2020tfl}.

\section{Software description}
\label{sec:descr}
The goal of cWB is to identify coherent GW transients on a network of GW 
detectors with minimal assumptions on signal morphology.
First, as shown in Fig. \ref{f:arch}, data streams from all detector are conditioned with a regression algorithm \cite{Tiwari:2015ofa} which identifies and removes persistent lines and noise artifacts; next, data are converted to the time-frequency (TF) domain with the Wilson-Daubechies-Meyer  (WDM) wavelet transform, which has very good localization properties, both in time and frequency \cite{Necula:2012zz}. 
The data are then whitened and those pixels whose energy is larger then a given threshold are retained for further analysis (see, e.g., Fig. \ref{f:TFexamples}).
The selected TF pixels from all detectors are combined in a constrained likelihood function that depends on the source sky position and 
takes into account the corresponding antenna patterns of the interferometers and time delays between interferometer pairs: after maximizing the constrained likelihood, 
a candidate event is identified when a specific measure of signal coherence, calculated on the ensemble of the selected TF pixels,   
exceeds a predetermined threshold \cite{Klimenko:2008fu,Klimenko:2005xv,Klimenko:2006rh}.

\subsection{Software Architecture}
\label{sec:arch}

 The core computational tasks are all performed by a specialized C++ library, the \textit{Wavelet Analysis Tool} (WAT) library, 
 and are embedded within the CERN ROOT data-analysis framework \cite{Brun:1997pa,Antcheva:2009zz}. ROOT classes are also used as building blocks for data processing (e.g. I/O), data analysis (e.g. post-production statistical analysis), and visualization. ROOT-derived classes and cWB classes are also accessible via CLING, the ROOT just-in-time interpreter \cite{Vasilev:2012ev}, and can be used with a C++ interactive shell or within ROOT C++ macros.
Python scripts are used for the most crucial tasks of cWB low-latency analysis, such as, for example, the trigger uploading to GraceDB (Gravitational-wave Candidate Event Database) \cite{gracedb}, which is a web service that provides early-warning information about candidate events.

cWB is built with the Frame Library \cite{framelib} to access data frames (the standard format for LIGO-Virgo data) and usually with support for skymap grids via HEALPix \cite{Gorski:2004by}, for the astronomical data format FITS \cite{CFITSIO} and for GW waveforms through the LALsuite library \cite{lalsuite}. 
 
In order to quickly adapt to the new needs that arise with the growing theoretical and experimental understanding of GWs, cWB is complemented by a large set of user-selectable options and by the possibility for the user to execute custom plugins which can be called at different stages of the pipeline. 

\subsection{Software Functionalities}
\label{sec:function}
\begin{figure}[h] 
	\centering
	\includegraphics[width=0.8\textwidth]{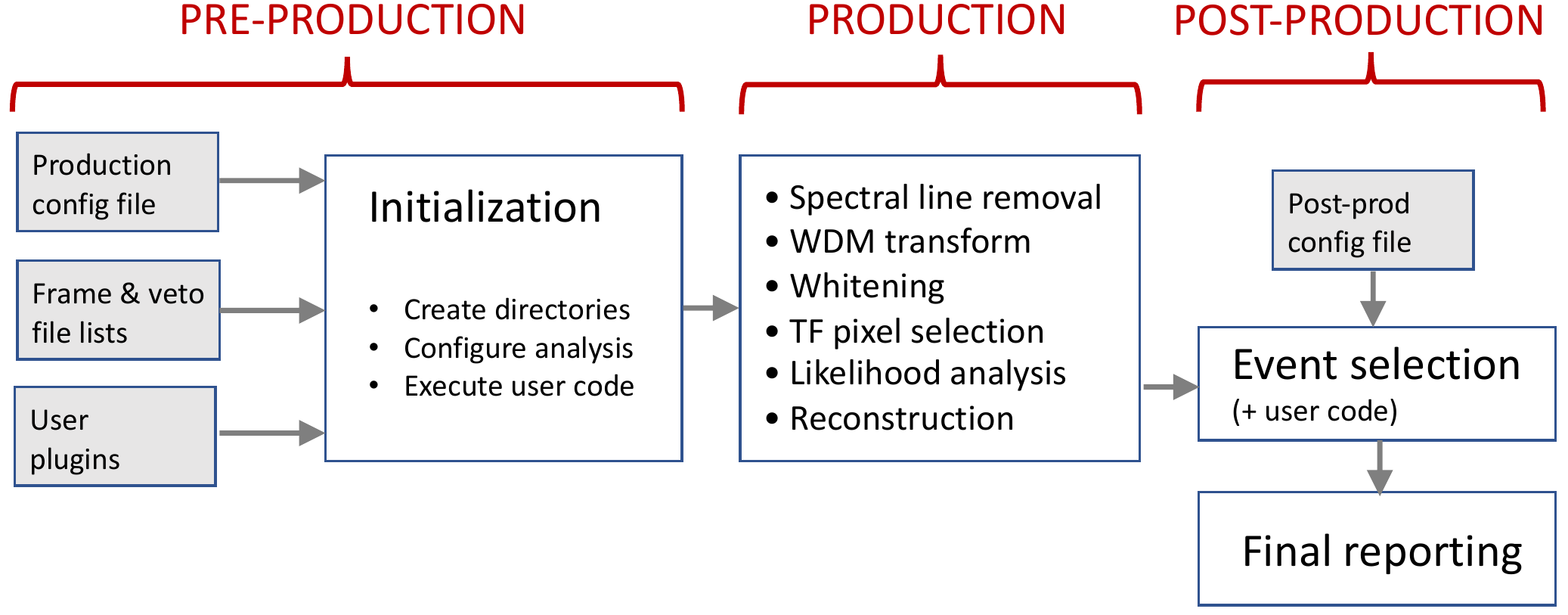} 
	\caption{Functional diagram of the cWB pipeline. White boxes show the processing steps, gray boxes show the external input. During PRE-PRODUCTION the pipeline initializes by reading the production configuration file which specifies the parameters that steer production. It also reads the \textit{frame files}, which contain strain and other data, and \textit{veto files} which mark those time intervals that must be excluded from analysis because of low data quality. User plugins that contain user-defined code to carry out targeted analyses are also loaded in this stage. Next, during PRODUCTION, the pipeline searches for triggers, and when one is found it reconstructs the gravitational waveform. In the POST-PRODUCTION stage, events are selected according to conditions specified in the initialization files and veto files, and finally, the pipeline outputs a  report in the form of a properly formatted web page.
	}
	\label{f:arch}
\end{figure}
The user can steer the pipeline behavior by setting parameter values in two distinct configuration files, one for the production stage and the other for the post-production stage, as indicated in the functional diagram in Fig. \ref{f:arch}. The production configuration file selects the interferometers (up to 8 detectors from a list of existing, possible future or custom interferometers) and the time segments to be included in the analysis, the parameters that regulate signal conditioning, etc. A full list of all production stage parameters is available in the online documentation of cWB \cite{cWB-params}. The production configuration file may also specify user-defined plugins, i.e. C++ code that can be called by cWB at different stages of the analysis. Using plugins, the user can customize the analysis without directly modifying cWB source code \cite{cwb-plugins}.
The post-production configuration file specifies further selections to be applied to web reports. The complete description of post-production parameter is also available in the online documentation \cite{cWB-pparams}.

Finally, cWB evaluates the accidental noise background by repeating each search several thousands of times on time-shifted data, and it injects specific waveforms in background data to estimate the statistical fluctuations of the reconstruction process.
For each reconstructed event, cWB estimates a large set of parameters and test statistics and can optionally produce a \textit{Coherent Event Display} (CED), the summary web page for the event. It consists of a number of sections, with each section showing plots on different aspects of the reconstruction \cite{ced}. 

All in all, cWB is quite efficient in carrying out its tasks. For example, a job that estimates the equivalent of 1 year of background noise for a BBH search on a two-detector network and runs on a single-core modern CPU takes approximately 6 hours. 
Parallelization is easy to achieve by splitting data into shorter time segments and by assigning them to separate jobs that replicate the same analysis.
Therefore, a larger analysis like a background estimate over thousands of years of time-lagged data can be completed in a matter of hours on the Caltech LIGO HPC cluster
\cite{Huerta_2017} where it can be run concurrently over thousands of cores. As far as memory allocation is concerned, the standard cWB setup requires about 1.5~GByte per core. A three-detector network brings about only a moderate increase in memory allocation, while runtime becomes approximately three times as large with respect to the two-detector network, because of the increased complexity of the analysis. 

\section{Illustrative Examples}
\label{sec:example}

The easiest way to test cWB functionalities is to install it as an image in a virtual environment: cWB images are available both for VirtualBox \cite{vbox} and Docker 
\cite{docker}. Full instructions on how to get the latest cWB images and run cWB are given in the cWB User Manual FAQs \cite{cwb-doc}

\subsection{The GW150914 example: cWB waveform reconstruction}
\label{GWOSC_example1}

By using the command {\tt cwb\_gwosc} \cite{cwb-gwosc}, it is straightforward to reproduce the full analysis of GW150914, the first detected GW event. The command loads GW data available from the Gravitational Wave Open Science Center (GWOSC) \cite{GWOS}, which includes the original raw data, the power spectral densities (PSDs), and the parameter-estimation (PE) posteriors samples for all GW detections. 

The command line to execute for the GW150914 event is 
{\small
\colorlet{shadecolor}{yellow!20}
\begin{shaded}
{\tt cwb\_gwosc GW=GW150914 {\color{green}all} }
\end{shaded}
}
which downloads GW data from GWOSC, sets up a working directory with all default settings, runs the analysis\footnote{Execution time is $\approx 2$ minutes on a commercial Intel i7 CPU laptop ($\approx +10\%$ on docker), excluding the initial GW data download.} and finally produces a CED \footnote{The CED can be visualized directly by replacing the command option {\tt  {\color{green}all}} with {\tt  {\color{green}xced}}. } \cite{cwb-ced}.
Fig. \ref{f:SIMexamples} (left) shows the multi-resolution TF map for the cWB reconstruction of GW150914. 

\subsubsection{cWB reconstruction on simulated data}
\label{GWOSC_example2}
\begin{figure*}[h]
\includegraphics[clip,width=0.49\textwidth]{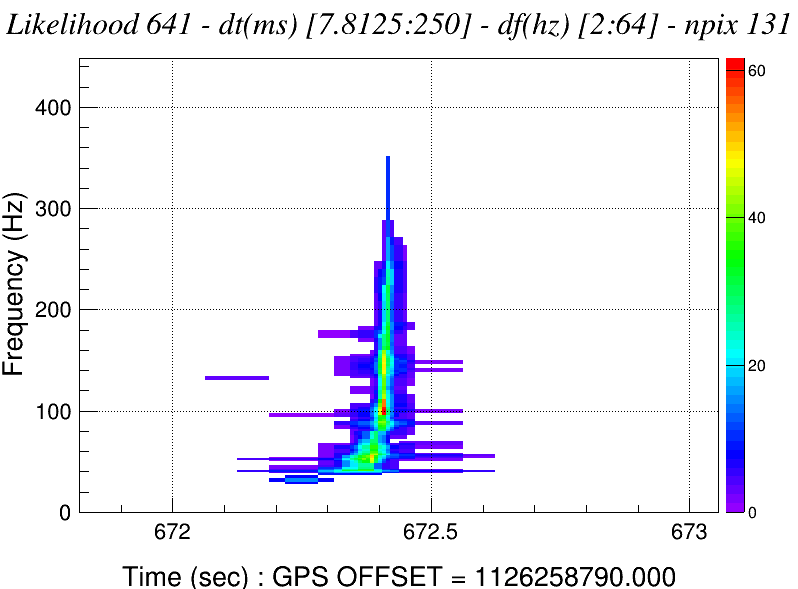}
	\hfil
	\includegraphics[clip,width=0.49\textwidth]{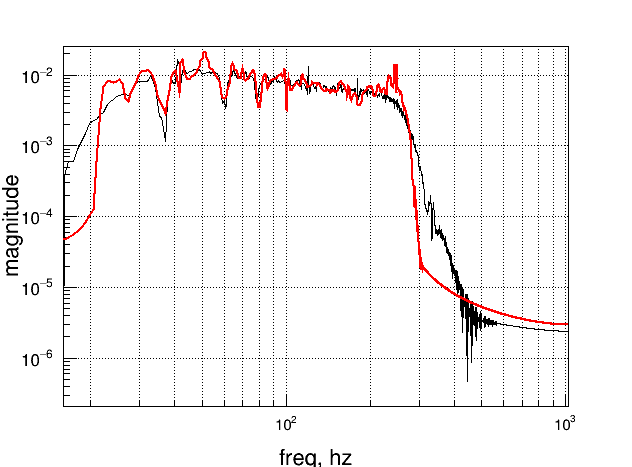}
	\caption{\label{f:SIMexamples} cWB waveform reconstruction of GW150914 as a color-coded TF map of the signal likelihood (left panel) and a comparison in the frequency domain (right panel) between a random posterior sample for GW150914 (black) and the corresponding cWB reconstruction (red).}
\end{figure*}

The posterior samples made available at GWOSC can be injected into simulated noise and reconstructed by cWB in order to estimate the variability of the reconstruction process for GW150914. The {\tt cwb\_gwosc} command is used to perform this analysis: 
{\small
	\colorlet{shadecolor}{yellow!20}
	\begin{shaded}
		{\tt cwb\_gwosc GW=GW150914 SIM=true {\color{green}all} }
	\end{shaded}
}
This instruction downloads the noise power spectral density (PSD) and a set of posterior samples for GW150914. Next, the PSDs are used to simulate a colored Gaussian stationary noise and a random sample waveform is added to the data. Finally, cWB runs as in the previous example and reports the results on a CED. Fig. \ref{f:SIMexamples} (right) shows a comparison between injected and reconstructed waveforms in the frequency domain.     
    
\subsubsection{cWB WDM transform on GW data and TF pixel selection}

The intermediate steps of the cWB analysis can be visualized with the {\tt cwb\_inet2G} command with plenty of options \cite{cwb-inet}. This is demonstrated for the pixel selection step of the example in Sec. \ref{GWOSC_example1} by running the following  commands:
{\small
\colorlet{shadecolor}{yellow!20}
\begin{shaded}
{\tt cwb\_inet2G config/user\_parameters CSTRAIN 1 {\color{green}'--tool wdm --ifo H1 --type white --draw true'} }

{\tt cwb\_inet2G config/user\_parameters SUPERCLUSTER 1 {\color{green}'--tool sparse --ifo H1 --type supercluster --draw true --mode 2'} }
\end{shaded}
}

\begin{figure*}[h]
\centering
	\includegraphics[width=\textwidth]{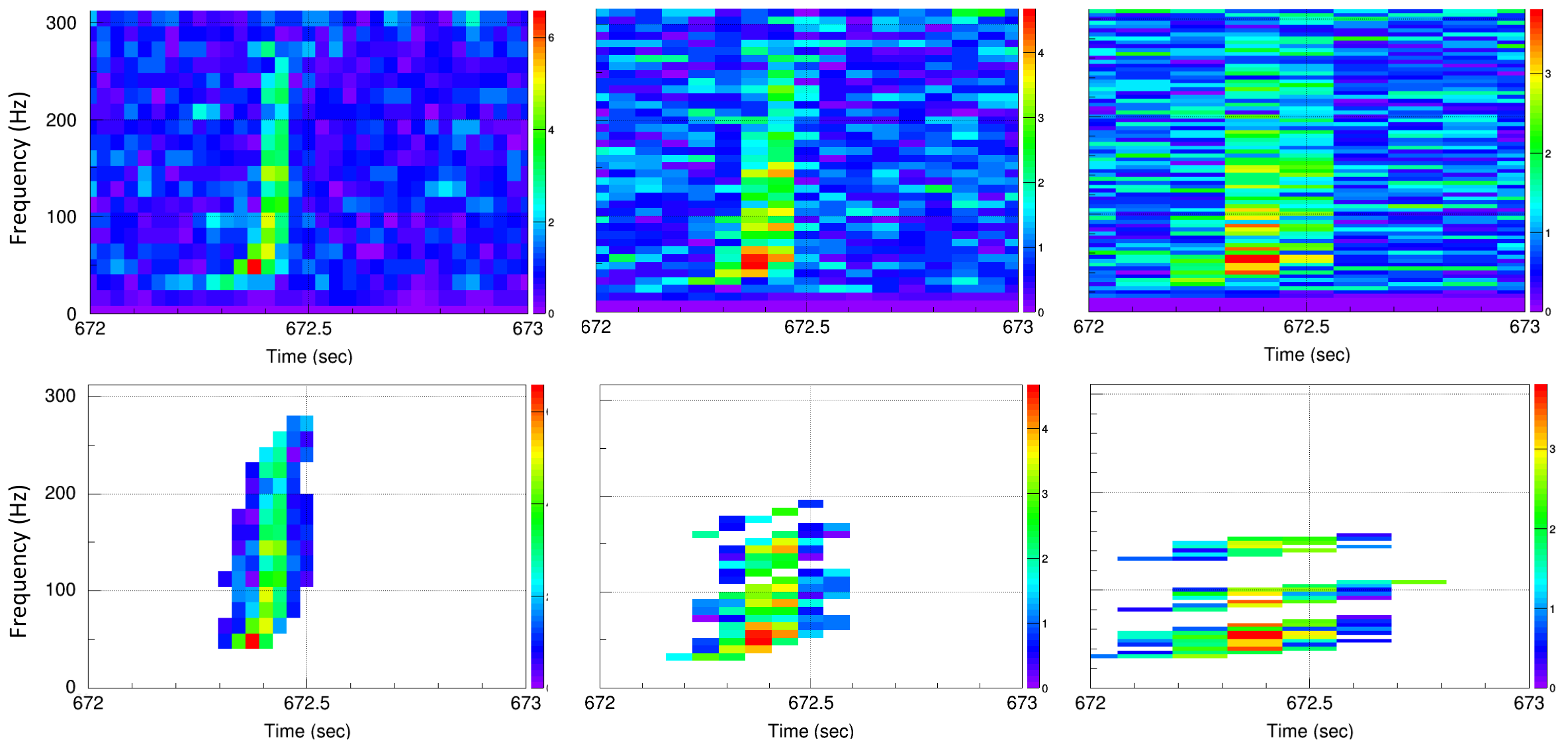}
	\caption{\label{f:TFexamples} Scalograms of Hanford data around GW150914: upper row, TF maps showing all data; lower row, TF maps showing only pixels retained for likelihood analysis. The columns correspond to different TF resolutions: left column, 1/32~s$\;\times\;$16~Hz; middle column: 1/16~s$\;\times\;$8~Hz; right column, 1/8~s$\;\times\;$4~Hz. }
\end{figure*}

Various plots are created as interactive objects and sorted within the ROOT browser (as described in the documentation \cite{cwb-doc}); among those some TF plots ( properly zoomed in time and frequency) have been selected to be shown in Fig. \ref{f:TFexamples}. The plots display the TF pixels from the WDM transforms for Hanford data at the time of GW150914 for three different TF resolutions, both before and after the pixel selection step. For more examples, see the cWB Manual \cite{cwb-doc}.

\section{Impact}
\label{sec:impact}
 
For more than a decade, coherent WaveBurst has been used for the analysis of GW data and for the production of scientific results, such as 
the all-sky searches for  burst signals targeting a wide class of generic bursts\cite{Abbott:2019prv,Abbott:2016ezn}. 
The cWB pipeline has also contributed to the detection and analysis of binary black hole events published in  the LIGO-Virgo Catalogs \cite{LIGOScientific:2018mvr,Abbott:2020niy} and to the LIGO-Virgo electromagnetic follow-up program \cite{LIGOScientific:2019gag}. 
Dedicated cWB searches for the intermediate-mass black hole binaries were also conducted, which set stringent limits approaching interesting astrophysical rates \cite{Salemi:2019ovz,Abbott:2017iws}.

Most importantly, cWB boosted exceptional discoveries, like the first detection of gravitational
waves from a merger of two black holes.
cWB was the first LIGO algorithm to report the event with low latency.  The results of the cWB analysis were published in the GW150914 discovery paper \cite{Abbott:2016blz}. cWB contribution has been acknowledged in the official description \cite{Nobel} of the 2017 Nobel prize in physics awarded to Rainer Weiss, Barry C. Barish and Kip S. Thorne. 
The broader impact of cWB has been highlighted also by Prof. Yves Meyer, the recipient of 2017 Abel Prize, who acknowledged the cWB wavelet analysis in his prize lecture “Detection of gravitational waves and time-frequency wavelets” at the University of Oslo on May 24, 2017 \cite{AbelLecture}.
More recently, cWB played a key role in the detection of the first intermediate-mass black hole \cite{Abbott:2020tfl}, as briefly mentioned in Sec.\ref{sec:intro}.

Although cWB has been designed with gravitational data analysis in mind, it integrates search\&reconstruction techniques for low SNR signals in the audio-band that may find useful applications in other fields, such as, for example, in the field of sound pattern recognition.

\section{Conclusions}
\label{sec:conc}

The cWB pipeline has some distinctive strengths: 
it is fast and efficient, it is highly flexible, and it is extensible -- both with specific plugins and, e.g. with the use of machine learning algorithms for recognition of patterns in the TF maps. It is ready to analyze data from larger (future) GW networks: networks composed of up to 8 detectors, chosen among the defaults or directly defined are allowed. These strengths may be crucial in view of the planned upgrades to GW observatories \cite{collaboration2020prospects}, which shall bring about increased sensitivities with correspondingly higher event rates: we expect cWB to play a key role in both low-latency and offline analysis.

\section{CRediT author statement}
M. Drago: Formal analysis, Software, Writing - Review \& Editing; V. Gayathri: Formal analysis; S. Klimenko: Conceptualization, Project administration, Supervision, Methodology, Software, Writing - Review \& Editing, C. Lazzaro: Formal analysis, E. Milotti: Methodology, Writing - Original Draft and Review \& Editing, G. Mitselmakher: Conceptualization; V. Necula: Methodology, Software; B. O'Brien: Formal analysis; G.A. Prodi: Supervision, Methodology, Writing - Review \& Editing; F. Salemi: Writing - Original Draft and Review \& Editing, Software, Methodology, Formal analysis; M. Szczepanczyk: Formal analysis; S. Tiwari: Formal analysis; V. Tiwari: Formal analysis, Methodology; G. Vedovato: Software, Methodology, Formal analysis, Data Curation, Visualization, Writing - Review \& Editing; I. Yakushin: Software, Formal analysis, Data Curation.

\section{Conflict of Interest}

No conflict of interest exists:
We wish to confirm that there are no known conflicts of interest associated with this publication and there has been no significant financial support for this work that could have influenced its outcome.

\section*{Acknowledgements}
\label{sec:ack}

cWB makes use of data, software and/or web tools obtained from the Gravitational Wave Open Science Center \cite{gwosc}, a service of LIGO Laboratory, the LIGO Scientific Collaboration and the Virgo Collaboration. LIGO is funded by the U.S. National Science Foundation. Virgo is funded by the French Centre National de Recherche Scientifique (CNRS), the Italian Istituto Nazionale della Fisica Nucleare (INFN) and the Dutch Nikhef, with contributions by Polish and Hungarian institutes. 
S.Tiwari is supported by University of Zurich Forschungskredit {Nr. FK-19-114} and Swiss National Science Foundation. 
Finally, we gratefully acknowledge the support of the
State of Niedersachsen/Germany, of the National Science Foundation (NSF)
and of the Istituto Nazionale di Fisica Nucleare (INFN) for provision of computational resources.

  \bibliographystyle{elsarticle-num} 
  \bibliography{references}

\end{document}